\pdfoutput=1
\documentclass[aps,pr,groupedaddress]{revtex4-2}

\usepackage{amsmath}
\usepackage{graphicx}
\usepackage{color}
\usepackage{bm}
\usepackage{graphicx}
\usepackage{amssymb}
\usepackage{tikz}
\usepackage{pgfplots}
\usepackage{soul,xcolor}
\usepackage{natbib}
\usepackage{natbib}
\usepackage{gensymb}
\usepackage{amsmath, bm}
\setstcolor{red}

\begin{document}

\title{Electron spin-vorticity coupling in low and high Reynolds number pipe flows} 

\author{Hamid Tabaei Kazerooni}
\email{hamid.kazerooni@tu-ilmenau.de}
\affiliation{Institute of Thermodynamics and Fluid Mechanics, Technische Universit\"at Ilmenau, 98684 Ilmenau, Germany}
\author{Alexander Thieme}
\email{alexander.thieme@tu-ilmenau.de}
\affiliation{Institute of Thermodynamics and Fluid Mechanics, Technische Universit\"at Ilmenau, 98684 Ilmenau, Germany}
\author{J\"org Schumacher}
\email{joerg.schumacher@tu-ilmenau.de}
\affiliation{Institute of Thermodynamics and Fluid Mechanics, Technische Universit\"at Ilmenau, 98684 Ilmenau, Germany}
\author{Christian Cierpka}
\email{christian.cierpka@tu-ilmenau.de}
\affiliation{Institute of Thermodynamics and Fluid Mechanics, Technische Universit\"at Ilmenau, 98684 Ilmenau, Germany}
\date{\today}

\begin{abstract}
Spin hydrodynamic coupling is a recently discovered method to directly generate electricity from an electrically conducting fluid flow in the absence of Lorentz forces. This method relies on a collective coupling of electron spins -- the internal quantum mechanical angular momentum of the electrons -- with the local vorticity of a fluid flow. In this work, we experimentally investigate the spin hydrodynamic coupling in circular and non-circular capillary pipe flows and extend a previously obtained range of Reynolds numbers to smaller and larger values, $20<Re<21,500$, using the conducting liquid metal alloy GaInSn as the working liquid. In particular, we provide experimental evidences for the linear dependence of the generated electrical voltage with respect to the bulk flow velocity in the laminar regime of the circular pipe flow as predicted by Matsuo \textit{et al.} [Phys. Rev. B. 96, 020401 (2017)]. Moreover, we show analytically that this behavior is universal in laminar regime regardless of the cross-sectional shape of the pipe. Finally, the proposed scaling law by Takahashi \textit{et al.} [Nat. Phys. 12, 52 (2016)] for the generated voltage in turbulent circular pipe flows is experimentally evaluated at Reynolds numbers higher than in previous studies. Our results verify the reliability of the proposed scaling law for Reynolds numbers up to $Re=21,500$ for which the flow is in a fully developed turbulent state.

\end{abstract}

\maketitle

\section{Introduction} \label{intro}
Fluid dynamical processes in connection with quantum many body phenomena are known from quantum turbulence at low temperatures close to absolute zero \cite{Vinen2007,fonda2019reconnection} and strongly correlated electron systems in graphene that resemble transport properties of a classical viscous fluid \cite{levitov2016electron} to mention two prominent examples for bosonic and fermionic systems, respectively. Besides the electric charge, an electron carries an intrinsic angular momentum known as electron spin or in short spin. Over the past thirty years, this fundamental quantum property of an electron spawned a multidisciplinary research field called spintronics analogous to electronics \cite{wolf2001spintronics}. The ultimate goal of spintornics is to exploit the electron spins along with their charges to fabricate more efficient, faster and smaller devices for data processing and storage. In this context, as for the charge current in conventional electronic devices, the generation and manipulation of the flow of electron spins, i.e. spin current, are the main tasks in spintronics \cite{vzutic2004spintronics, hirohata2014future}. However, these are very challenging to realize as, unlike the charge current, the spin current is not a conserved quantity. Nevertheless, different approaches are proposed to drive the spin current mainly based on angular momentum exchange between the electron spin and other physical entities such as light polarization \cite{ando2010direct} and magnetization \cite{sandweg2011spin}. \citet{uchida2008observation} showed that a temperature gradient can also induce a spin current in a metallic magnet, i.e. the spin Seebeck effect. 

Recently, \citet{takahashi2016spin} investigated the spin current generation through a direct coupling between electron spins and the microscopic rotation of a material system as the most well-known form of angular momentum \cite{matsuo2013mechanical}. In particular, they utilized the vorticity generated by a turbulent flow of liquid metals in narrow circular pipes as a source of mechanical rotation. In a fluid flow, vorticity is a measure of the local rotation of a fluid element and it is mathematically defined as ${\bm \omega}={\bm \nabla} \times {\bm u}$, where ${\bm u}=(u_x,u_y,u_z)$ is the three-dimensional flow velocity field. \citet{takahashi2016spin} argued that the electron spins are polarized in the direction of the flow vorticity due the spin-mechanical angular momentum coupling. Moreover, they showed that the gradient of the vorticity results in a spin voltage gradient and consequently a spin current in the direction of the vorticity gradient. It has been found that a spin current induces an electrical current transversal to the spin current direction due to the spin-orbit interaction \cite{sinova2015spin}. This phenomenon is known as the inverse spin Hall effect (ISHE) (see \citet{saitoh2006conversion}) and employed by \citet{takahashi2016spin} to detect the induced charge current as a faint voltage $V_{\mathrm{ISHE}}$ signal in the flow direction of a turbulent pipe flow where the mean vorticity changes along the radial direction \cite{white2015fluid}. Based on a theoretical analysis, they also predicted that the voltage $V_{\mathrm{ISHE}}$ which is measured along the pipe flow direction should be mainly proportional to the square of the turbulent friction velocity $u_{\tau}$ in a circular pipe flow and its magnitude is of the order of nano-volt. Their experimental results are in very good agreement with their theoretical prediction using mercury (Hg) and the eutectic alloy made of galium, indium, and tin (GaInSn) as working liquids. Later on, \citet{matsuo2017theory} showed theoretically that the voltage $V_{\mathrm{ISHE}}$ increases linearly with respect to the bulk flow velocity  $U_b$ in laminar circular pipe flows. However, they did not provide any experimental evidence for this claim. It is one of the main motivations of the present study to fill this gap and to verify the voltage generation in an independent experimental setup which we will present in the following.

To do so, we have designed and developed an experimental apparatus which allows us to measure the generated voltage $V_{\mathrm{ISHE}}$ in both laminar and turbulent regimes and thus to extend the range of previously obtained flow Reynolds numbers. Our measurements are performed using circular capillary tubes with diameters $D$ ranging from 0.1 mm to 1.2 mm for a wide range of Reynolds numbers $Re=U_bD/\nu$ from 20 to 21,500 where $U_b$ and $\nu=3.3\times10^{-7}$ m\textsuperscript{2}s\textsuperscript{-1} are the bulk flow velocity and the kinematic viscosity of GaInSn as the working liquid, respectively. The GaInSn eutectic alloy is used due to its low kinematic viscosity, low melting temperature (10.5 $^{\circ}$C) and, importantly, its low toxicity. 

The present experiments extend the previous measurements of \citet{takahashi2016spin} in two directions. (1) Our results show that the generated voltage $V_{\mathrm{ISHE}}^{\mathrm{lam}}$, as predicted by \citet{matsuo2017theory}, is linearly proportional to the bulk flow velocity $U_b$ in laminar regime. To better understand the role of the capillary tube cross sectional shape on the spin hydrodynamic coupling, measurements were also carried out in a rectangular capillary tube and laminar flow regime. (2) In the turbulent regime, we investigate the generated voltage $V_{\mathrm{ISHE}}^{\mathrm{turb}}$ at four times higher Reynolds numbers than previous studies which were limited to $Re<5000$ using GaInSn as the working liquid. This brings us into the fully developed regime of pipe flow turbulence. Our work is an independent confirmation of the measurements of \citet{takahashi2016spin}. We find a very good agreement with their data for the same range of Reynolds numbers and GaInSn as the working liquid. Moreover, the results of the present study give strong evidence for an extended validity of the universal scaling law proposed by \citet{takahashi2016spin} for flows at high Reynolds numbers.
\section{Experimental setup}
\begin{figure}[t]
	\centering
	\includegraphics[scale=0.5,trim={1cm 0.5cm 7cm 0cm},clip]{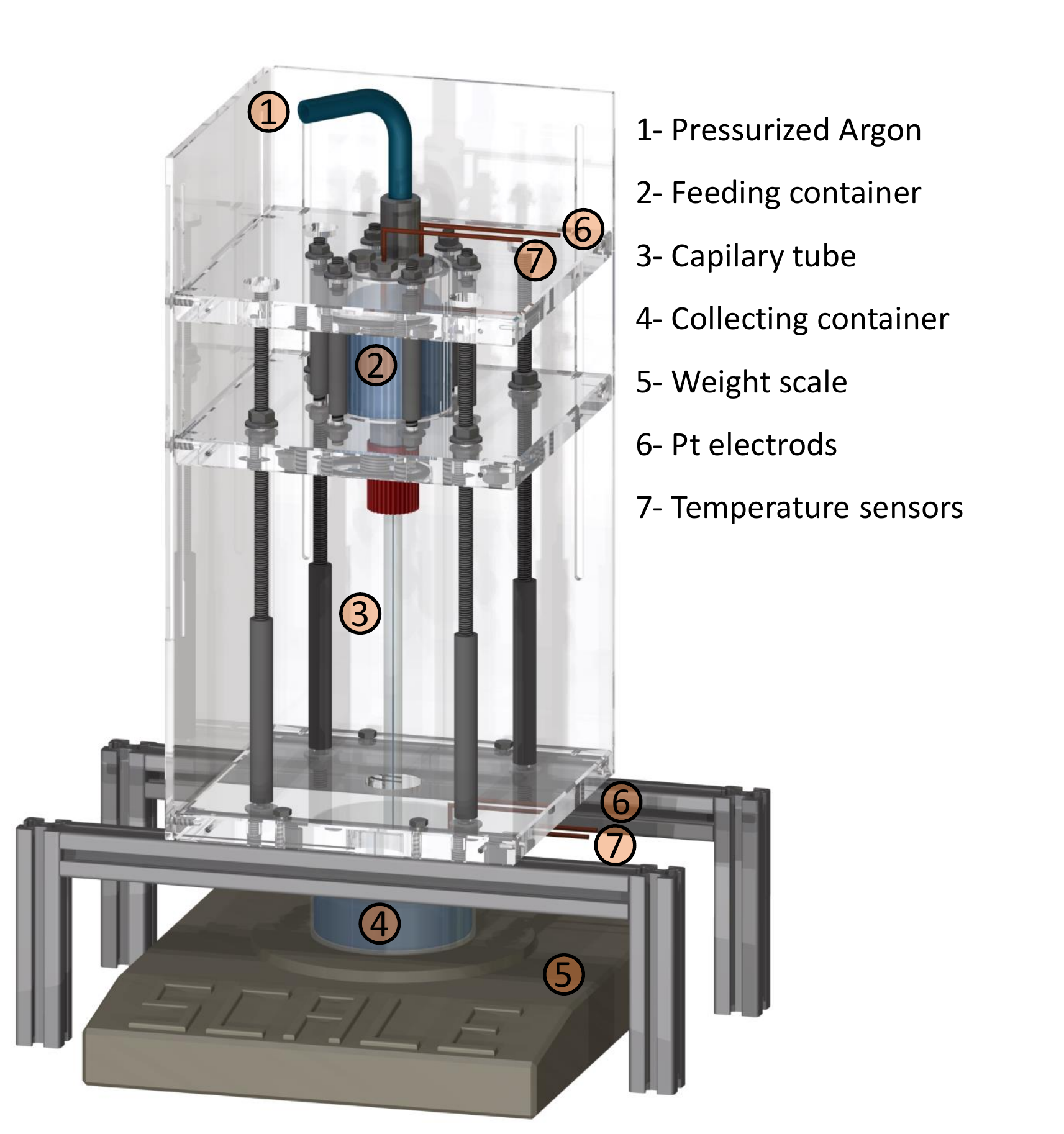}
	\caption{Schematic illustration of the employed experimental set-up for measuring the inverse spin hall voltage $V_{\rm ISHE}$ generated by laminar and turbulent flows in different narrow capillary tubes.}
	\label{Exp_Setup}
\end{figure}

The experimental set-up is designed to investigate the spin hydrodynamic coupling and measuring the regarding generated electrical voltage $V_{\mathrm{ISHE}}$ produced by laminar and turbulent flows of GaInSn in different capillary tubes. Figure.~\ref{Exp_Setup} shows a schematic illustration of the apparatus which comprises a pressurized feeding vessel (2) and a collecting container (4) both made of plexiglass, a capillary tube (3), a weight scale (5) for estimating the liquid metal flow rate and several sensors for monitoring the temperature (7) and measuring the longitudinal generated voltage $V_{\mathrm{ISHE}}$ (6). To conduct experiments, the liquid metal is loaded into the top sealed vessel with a total volume of 63 ml. The flow is driven to the capillary tube using pressurized Argon gas. A precise weight scale (Sartorius AG) is used to measure the weight change of the collecting container in order to accurately calculate the flow rate and the corresponding bulk flow velocity $U_b$ based on GaInSn density $\rho=6330$ kg/m\textsuperscript{3} and the measurement time. Table~\ref{table:1} presents all physical properties of GaInSn alloy used in this study and taken form \citet{plevachuk2014} and \citet{Sven}. It is known that the surface of GaInSn can be easily oxidized in the presence of air. This effect was minimized using the noble gas Argon to pressurize the liquid metal and generate the flow. Nevertheless, a layer of oxide was formed on the surface of the liquid metal in the top and bottom vessels. It is worth mentioning that these oxide layers remained unperturbed at the surface and they were not mixed with the whole liquid during the experiment. 

Capillary tubes with different diameters, lengths and cross-sections can be easily mounted via different adapters to the system. The majority of measurements were carried out using capillary tubes made of borosilicate glass (Hilgenberg GmbH) with circular cross-section and different diameters of $D$= 0.1, 0.4, 0.8 and 1.2 mm to cover a wide range of Reynolds numbers $Re$. In order to investigate the effect of the capillary tube cross-sectional shape on the spin hydrodynamic coupling in laminar regime, additional measurements were performed using a rectangular capillary tube with a sectional dimension of $W\times H= 0.334\times0.149~\mathrm{mm^2}$ and a hydraulic diameter of $D_{h_{Rec}}=2WH/(W+H)=0.2$~$\mathrm{mm}$. For the sake of comparability, the length of all capillaries was chosen to be $L=200$ mm.   
\begin{table}
\centering
\begin{tabular}{ll}
\hline\hline
Physical quantity        & Value (SI units) \\ 
\hline
Mass density              & $\rho = 6330 \;\mathrm{kg}/ \mathrm{m^3}$ \\
Kinematic Viscosity    & $\nu = 3.30\times 10^{-7} \;\mathrm{m^2}/\mathrm{s}$ \\
Thermal Diffusivisity    & $\kappa = 1.1\times 10^{-5} \;\mathrm{m^2}/\mathrm{s}$ \\
Magnetic Diffusivity    & $\beta = 0.24 \;\mathrm{m^2}/\mathrm{s}$ \\
Electrical conductivity & $\sigma_{0} = 3.26\times 10^6 \;(\Omega \mathrm{m})^{-1}$ \\
\hline\hline
\end{tabular}
\caption{List of the material properties of GaInSn at $26^{\circ}$C taken from Ref.~\cite{plevachuk2014}. The value of the specific heat at constant pressure $c_p$, that is required to calculate $\kappa$, is obtained by extrapolation from data at higher temperatures.}
\label{table:1} 
\end{table}
The generated voltage $V_{\mathrm{ISHE}}$ was measured by means of a nanovoltmeter (KEITHLEY 2182A) and two very thin wires ($d=0.4$ mm) made of Platinum (Pt) implemented in the top and bottom vessels as electrodes. The liquid GaInSn in the top vessel is connected to ground to prevent the accumulation of static electric charge. The temperature of the vessels was also monitored using calibrated PT100 resistance thermometers with an uncertainty of $\pm0.1 ^{\circ}$C. The whole apparatus was placed inside a thermally controlled chamber and the system was operated for a few hours before each set of experiments to reach stable conditions at $26^{\circ}$C. Nevertheless, temperature differences may occur due to the Argon gas compression and expansion in the set-up. Considering the upper vessel height $L=50$ mm and the thermal diffusivity $\kappa$ of GaInSn from Table I, the diffusion time of heat, $\tau=L^2/\kappa$, in the upper vessel can be roughly estimated to be about 200 seconds. To minimize thermal effects in the system, measurements were only carried out with a fully filled vessel and for a time duration of 10 seconds. Data acquisition was carried out using National Instruments LabVIEW\textsuperscript{TM} software package. The sampling frequency is set to 2 Hz corresponding to a maximum resolution of 1 nV for recording the generated voltage $V_{\mathrm{ISHE}}$ via the nanovoltmeter.
  
\section{Theoretical aspects} \label{Theory}

\begin{figure}[t]
	\centering
	\includegraphics[scale=0.37,trim={0cm 0cm 0cm  0cm},clip]{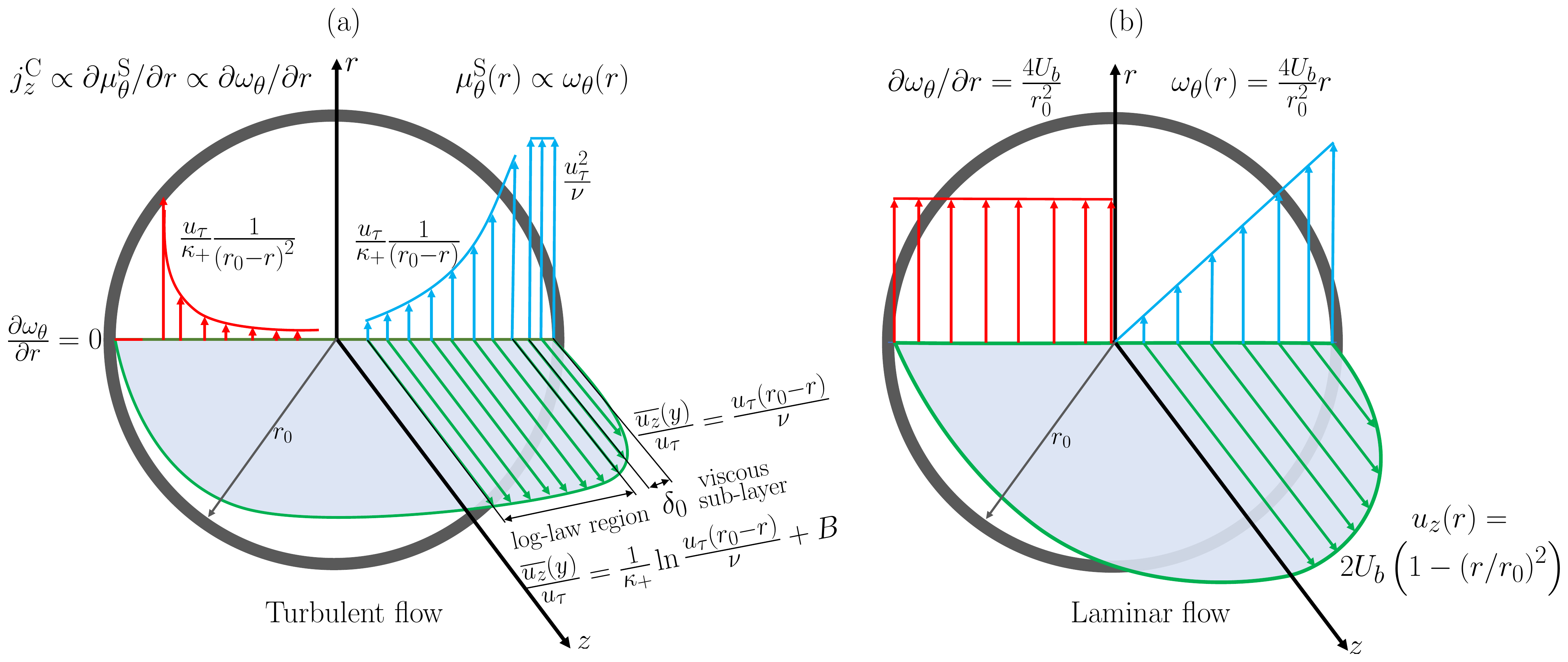}
	\caption{(a) Turbulent circular pipe flow mean velocity (green profile), mean vorticity field (blue profile) and the mean radial vorticity gradient (red profile). The electron spins are aligned along the mean vorticity direction. The spin voltage is proportional to the mean flow vorticity $\mu^{\mathrm{S}}_{\theta}(r) \propto \omega_{\theta}(r)$. The radial gradient of the spin voltage due to the change of the flow vorticity along the pipe radius induces charge current in the streamwise direction based on the inverse spin hall effect (ISHE) $j_{z}^{\mathrm{C}} \propto \partial\mu_{\theta}^{\mathrm{S}}/ \partial{r} \propto \partial\omega_{\theta}/ \partial{r}$. The gap between the viscous sub-layer and the log-law region which is called buffer layer is neglected by \citet{takahashi2016spin} in their analysis. (b) Parabolic velocity profile (green) of a laminar circular pipe flow and the corresponding vorticity field (blue profile) and the constant radial vorticity gradient (red profile). Vectors in these figures are only for illustration purposes and they are not scaled based on the magnitude of each quantity.}
	\label{Fig_Theory}
\end{figure}

As mentioned before, \citet{takahashi2016spin} were the first ones to discover and introduce the concept of the spin hydrodynamic generation. Besides from the experimental evidences, they have also provided a theoretical framework for explaining this phenomenon. In short, based on the angular-momentum conservation law in a fluid flow and the spin diffusion equation, they proposed an equation to describe the relationship between the vectorial spin voltage $\boldsymbol{\mu}^{\mathrm{S}}$ and the flow vorticity $\boldsymbol{\omega}$ vector. This extension of the Valet-Fert equation for spin diffusion \cite{valet1993theory} reads as follows:
\begin{equation}
\label{main_spv}
\nabla^{2} \boldsymbol{\mu}^{\mathrm{S}}=\frac{1}{\lambda^{2}} \boldsymbol{\mu}^{\mathrm{S}}-\frac{4 e^{2} \xi}{\sigma_{0} \hbar}  ~\boldsymbol{\omega}
\end{equation}
where $\lambda$, $e$, $\hbar$ and $\sigma_{0}$ are the spin diffusion length, the elementary charge, the reduced Plank constant and the electrical conductivity of the liquid metal, respectively. \citet{takahashi2016spin} defined $\xi$ as a parameter which represents the angular momentum transfer from the fluid into electron spins. They also argued that $\xi$ is different for laminar $\xi_{\mathrm{lam}}$ and turbulent $\xi_{\mathrm{turb}}$ flows. Note that the vector notation in eq.~(\ref{main_spv}) indicates the polarization direction of the spin voltage $\boldsymbol{\mu}^{\mathrm{S}}$ which corresponds to the flow vorticity direction. For more details on the derivation of eq.~(\ref{main_spv}), the reader is referred to \citet{takahashi2016spin}. It is also shown that the induced spin current $\mathbf{j}^{\mathrm{S}}$ and the electrical field $\mathbf{E}_{\mathrm{ISHE}}$ due to the ISHE can be described as:

\begin{equation}
\label{ele_field}
\mathbf{E}_{\mathrm{ISHE}}=-\frac{2|e|}{\sigma_{0} \hbar} \theta_{\mathrm{SHE}}~\mathbf{j}^{\mathrm{S}} \times \boldsymbol{\sigma}
\end{equation}
where $\theta_{\mathrm{SHE}}$ and $\boldsymbol{\sigma}$ are the spin Hall angle and the spin-polarization vector \cite{saitoh2006conversion}. Hence, the charge current can be written as $j_{i}^{\mathrm{C}}=-\frac{2|e|}{\hbar} \theta_{\mathrm{SHE}} \epsilon_{i j k} j_{j k}^{\mathrm{S}}$ using the Levi-Civita symbol $\epsilon_{i j k}$ in tensor notation. It is well-known that the average streamwise velocity profile in a wall-bounded pressure-driven turbulent flow close to the wall in the viscous sub-layer $\delta_{0}$ is a linear function of the wall-normal direction $y$ reads as:
\begin{equation}
\label{Lam_veloctiy}
\frac{\overline{u_{z}}(y)}{u_{\tau}}=\frac{yu_{\tau}}{\nu}
\end{equation}
where $u_{\tau}=\sqrt{\tau_{w}/\rho}$ is the turbulent friction velocity considering $\tau_{w}$ and $\rho$ as the wall-shear stress and the fluid density, respectively. It is also well-accepted that the log-law describes the velocity profile away from the viscous sub-layer along the wall normal direction $y$ in canonical flows as follows \cite{pope2001turbulent}:
\begin{equation}
\label{Turb_veloctiy}
\frac{\overline{u_{z}}(y)}{u_{\tau}}= \frac{1}{\kappa_+} \ln \left(\frac{yu_{\tau}}{\nu}\right)+B
\end{equation}
where $\kappa_+$ and $B$ are the von Kármán constant and the additive coefficient, respectively. \citet{takahashi2016spin} employed the above velocity profile considering $\kappa_+ \approx 0.41$ and $B \approx 5.5$ to calculate the vorticity field in a turbulent pipe flow with a circular cross-section in cylindrical coordinates. 

Figure.~\ref{Fig_Theory}(a) shows the mean flow velocity profile, the corresponding mean vorticity field and its gradient for a turbulent circular pipe flow in cylindrical coordinates with $z$ being the streamwise and $r$ the wall-normal direction, respectively. Clearly, the mean vorticity field $\boldsymbol{\omega}$ of a circular pipe flow consists of only one component in azimuthal $\theta$ direction $\omega_{\theta}(r)=-\partial_{r} \overline{u_{z}}(r)$. Based on eq.(\ref{ele_field}), the gradient of the vorticity and consequently the spin voltage in the radial direction $\partial\mu_{\theta}^{S}(r)/\partial{r} \propto \partial{\omega_{\theta}(r)}/\partial{r}$ is responsible for inducing the charge current along the streamwise direction $j_{z}^{\mathrm{C}}$. Hence, the mean vorticity in the viscous sub-layer $\omega_{\theta}(r)={u_{\tau}^{2}}/{\nu}$ does not contribute to the spin hydrodynamic generation as its radial gradient is zero. However, the radial vorticity gradient in the log-law region is not constant and can be expressed as:
\begin{equation}
\label{Turb_vorticity}
\frac{\partial{\omega_{\theta}(r)}}{\partial {r}}  =\frac{u_{\tau}}{\kappa_+} \frac{1}{\left(r_{0}-r \right)^{2}}
\end{equation}

Therefore, the radial vorticity gradient in the log-law region, especially close to the wall where it is more significant, is the main source of the induced spin and charge current in the same $r$ and transversal $z$ direction, respectively (see Fig.~\ref{Fig_Theory}(a)). Note that here the wall-normal direction in a Cartesian coordinate is transformed into the cylindrical one using $y=r-r_0$ where $r_0$ is the pipe radius. \citet{takahashi2016spin} employed eq.~(\ref{Turb_vorticity}) and analytically solved eqs.~(\ref{main_spv}) and (\ref{ele_field}) for a turbulent pipe flow with a circular cross-section. For the sake of simplicity, they neglected the buffer layer between the viscous sub-layer and the log-law region. Based on their analysis, \citet{takahashi2016spin} proposed a universal scaling law for the inverse spin Hall voltage $V_{\mathrm{ISHE}}^{\mathrm{turb}}$ generated in a turbulent circular pipe flow as a nonlinear function of the turbulent friction velocity $u_{\tau}$:

\begin{equation}
\label{uni_sc_turb}
\frac{r_{0}^{3}~V_{\mathrm{ISHE}}^{\mathrm{turb}}}{L}= \frac{4|e|}{ \hbar} \cdot \frac{\theta_{\mathrm{SHE}} \lambda^{2} \xi_{\mathrm{turb}}}{\sigma_{0}}\cdot \frac{1}{\kappa_+}\left[\frac{\left(u_{\tau} r_{0}\right)^{2}}{\nu Re_{\tau}^{\delta_{0}}}-\left(u_{\tau} r_{0}\right) \ln \frac{\left(u_{\tau} r_{0}\right)}{\nu Re_{\tau}^{\delta_{0}}}-\left(u_{\tau} r_{0}\right)\right]
\end{equation}
where $Re_{\tau}^{\delta_{0}}=u_{\tau}\delta_{0}/\nu \approx 11.6$ is the friction Reynolds number based on the viscous sub-layer thickness $\delta_{0}$. Note that this value for $Re_{\tau}^{\delta_{0}}$ can be achieved when one considers the viscous sub-layer thickness as the height, $y=r_0-r$, where the logarithmic velocity profile (see eq.~(\ref{Turb_veloctiy})) intersects with the linear velocity profile of the viscous sub-layer (see eq.~(\ref{Lam_veloctiy})). A series of experimental measurements by \citet{takahashi2016spin} showed the reliability of the proposed scaling law in a range of Reynolds number $Re$ from 4,000 to 10,000 using Hg as the working liquid.

The same procedure described above was conducted by \citet{matsuo2017theory} to obtain a universal scaling for the generated voltage $V_{\mathrm{ISHE}}$ in a laminar circular pipe flow with a parabolic velocity profile as:
\begin{equation}
\label{Lam_Veloc}
u_{z}(r)=2 U_{b}\left[1-\left(\frac{r}{r_0}\right)^{2}\right]
\end{equation}

As shown in Fig.~\ref{Fig_Theory}(b), the vorticity field has again one component only, $\omega_{\theta}(r)=4U_{b}r/r_0^2$ in azimuthal direction $\theta$ and its gradient in the radial direction $r$ is constant $\partial_{r} \omega_{\theta}(r)=4U_{b}/r_0^2$. The solution of eqs.~(\ref{main_spv}) and (\ref{ele_field}) for a laminar pipe flow indicates that the generated voltage $V_{\mathrm{ISHE}}^{\mathrm{lam}}$ is a linear function of the bulk flow velocity $U_b$ reads as:
\begin{equation}
\label{uni_sc_lam}
\frac{r_{0}^{3}~V_{\mathrm{ISHE}}^{\mathrm{lam}}}{L}= \frac{8|e|}{\hbar} \cdot \frac{\theta_{\mathrm{SHE}} \lambda^{2} \xi_{\mathrm{lam}}}{\sigma_{0}} \cdot (U_b~r_0)
\end{equation}
The validity of eq.~(\ref{uni_sc_lam}), which has not yet been proven experimentally, will be elaborated in the next section together with the results for turbulent flows at moderately high Reynolds numbers. 
\section{Measurement results}
\subsection{Low Reynolds number laminar flows}
In this section, we first present and discuss the obtained experimental results on the spin hydrodynamic coupling at low Reynolds numbers ($20<Re<1700$) where the flow is assumed to be fully laminar. Regardless of the flow regime, the voltage signal measurement was carried out for a total measurement time of 10 seconds to avoid possible thermoelectric effects. Indeed, despite a moderately high relative Seebeck coefficient of GaInSn and Pt electrodes $S_{\mathrm{GaInSn}-\mathrm{Pt}}=+ 4.48~\mu \mathrm{V} \mathrm{K}^{-1}$, no significant influence of the Seebeck effect on the measured voltage $V_{\mathrm{ISHE}}$ is observed as the temperature of the top and bottom vessels remain constant during this short period of time. 

\begin{figure}[h]
	\centering
	\includegraphics[width=1\textwidth]{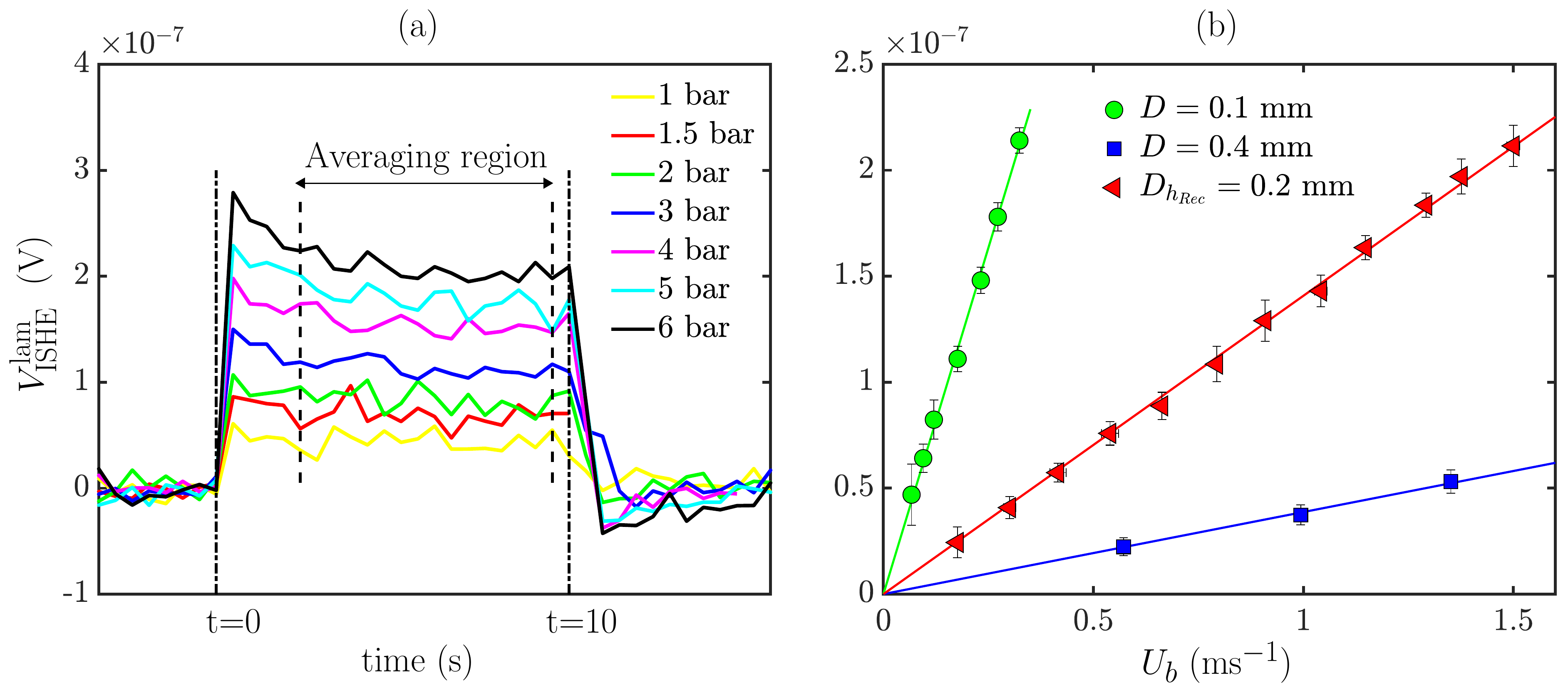}
	\caption{The inverse spin hall voltage $V_{\mathrm{ISHE}}^{\mathrm{lam}}$ generated by laminar flows in different capillary tubes with the same length of $L=200$ mm. (a) Time evolution of the voltage signal $V_{\mathrm{ISHE}}^{\mathrm{lam}}$ in a circular capillary tube with $D=$ 0.4 mm where the flow is maintained for 10 seconds. The region between the dashed lines, where the flow reaches steady state, is used to estimate the mean value of the generated voltage. The whole measurement interval of 10 seconds is indicated by the dash-dotted lines. (b) The generated voltage $V_{\mathrm{ISHE}}^{\mathrm{lam}}$ with respect to the bulk flow velocity $U_b$ in circular and a rectangular capillary tubes. Error bars show the standard deviation of at least five independent measurements.}
	\label{V_evo_t}
\end{figure}

Two sets of measurements were performed using narrow circular capillary tubes with diameters of $D=0.1$ mm and 0.4 mm to ensure that laminar flow could be established. Figure.~\ref{V_evo_t}(a) shows, as an example, the time evolution of the voltage $V_{\mathrm{ISHE}}^{\mathrm{lam}}$ signal for different imposed pressure or pressure differences ranging from 1 bar to 6 bar. The flow is generated at $t=0$ and is ceased after 10 seconds. No voltage $V_{\mathrm{ISHE}}^{\mathrm{lam}}\approx 0$ is recorded until the liquid metal is driven into the capillary. As soon as the flow starts, a sharp jump is observed in the voltage signal as clearly can be seen from Fig.~\ref{V_evo_t}(a). Note that the voltage signal is averaged over a shorter period time where the flow is steady and non-accelerating (indicated by the vertical dashed lines). Moreover, all the reported voltages $V_{\mathrm{ISHE}}$ in the present study are the results of at least five individual measurements at each pressure. However, for clarity, only one signal, as an example, is plotted for each pressure in Fig.~\ref{V_evo_t}(a) where the generated voltage $V_{\mathrm{ISHE}}^{\mathrm{lam}}$ is obviously increased with respect to the imposed pressure or the bulk flow velocity $U_b$. As said before, additional measurements in the laminar regime were carried out using a narrow capillary with a rectangular cross-section to investigate the effect of the tube geometry on the spin hydrodynamic generation. Figure~\ref{V_evo_t}(b) shows all the measured voltages for laminar cases with respect to the bulk flow velocity $U_b$ where the minimum and maximum Reynolds numbers $Re$ are almost 20 and 1700 for circular capillaries of $D=0.1$ mm and 0.4 mm, respectively. 

The results presented in Fig.~\ref{V_evo_t}(b) indicate the linear evolution of the generated voltage $V_{\mathrm{ISHE}}^{\mathrm{lam}}$ with respect to the bulk flow velocity $U_b$ for all investigated cases. Such a behavior was predicted by \citet{matsuo2017theory} for a laminar pipe flow with a circular cross-section. However, it is interesting to see the same behavior in a  rectangular capillary tube with an inhomogeneous distribution of the vorticity field across the capillary cross-section. This, of course, can be explained based on the previously mentioned relationship between the charge and spin current $j_{i}^{\mathrm{C}}=-\frac{2|e|}{\hbar} \theta_{\mathrm{SHE}} \epsilon_{i j k} j_{j k}^{\mathrm{S}}$ where $j_{j k}^{\mathrm{S}}$ is a second-rank tensor with nine components. Given this relationship, the longitudinal component of the induced charge current due to the ISHE in a Cartesian coordinate $j_{z}^{\mathrm{C}}$ should be only a result of the in-plane spin current $j^{\mathrm{S}}_{x-y}$, i.e.  $j_{z}^{\mathrm{C}}\propto (j_{xy}^{\mathrm{S}}-j_{yx}^{\mathrm{S}})$. Based on \citet{takahashi2008spin}, the spin current and consequently the charge current can be expressed as a spatial gradient of the spin voltage $j_{z}^{\mathrm{C}}\propto (\partial\mu_{y}^{S}/\partial{x}-\partial\mu_{x}^{S}/\partial{y})$. Using a polar coordinate system, $j_{z}^{\mathrm{C}}$ is reduced to $j_{z}^{\mathrm{C}}\propto j_{r\theta}^{\mathrm{S}} \propto \partial\mu_{\theta}^{S}/\partial{r}$ for the flow in a circular pipe with only one azimuthal in-plane ($r-\theta$) vorticity component $\omega_{\theta}(r)$. This allowed \citet{matsuo2017theory} to analytically solve eqs.~(\ref{main_spv}) and (\ref{ele_field}) and proposing eq.~(\ref{uni_sc_lam}) as a universal scaling law. In a rectangular pipe flow, however, given the presence of two different in-plane vorticity components of $\omega_{x}(x,y)$ and $\omega_{y}(x,y)$, the calculation becomes more complicated (as it involves a series expansion of the solution) and will be reported elsewhere. Nonetheless, it is known that the in-plane vorticity components in a laminar pressure-driven flow is linearly proportional to the bulk flow velocity $\omega_{x-y}=U_bf(x,y)$ regardless of the tube cross-sectional shape which its influence represented here by $f(x,y)$ as a 2D spatial function. This implies that whatever the solution of eq.~(\ref{main_spv}) for the spin voltage field $\mu_{x-y}^{\mathrm{S}}$ turns out to be, it should be again a linear function of the bulk flow velocity given the vorticity field $\omega_{x-y}=U_bf(x,y)$ as a source term on the right hand side of eq.~(\ref{main_spv}). Hence, as $j_{z}^{\mathrm{C}} \propto \nabla \mu \propto \nabla \omega$, it can be argued that the generated voltage in laminar regime $V_{\mathrm{ISHE}}^{\mathrm{lam}}$ is linearly proportional to the bulk flow velocity regardless of the tube cross-sectional shape.  
\begin{figure}[h!]
	\centering
	\includegraphics[width=1\textwidth]{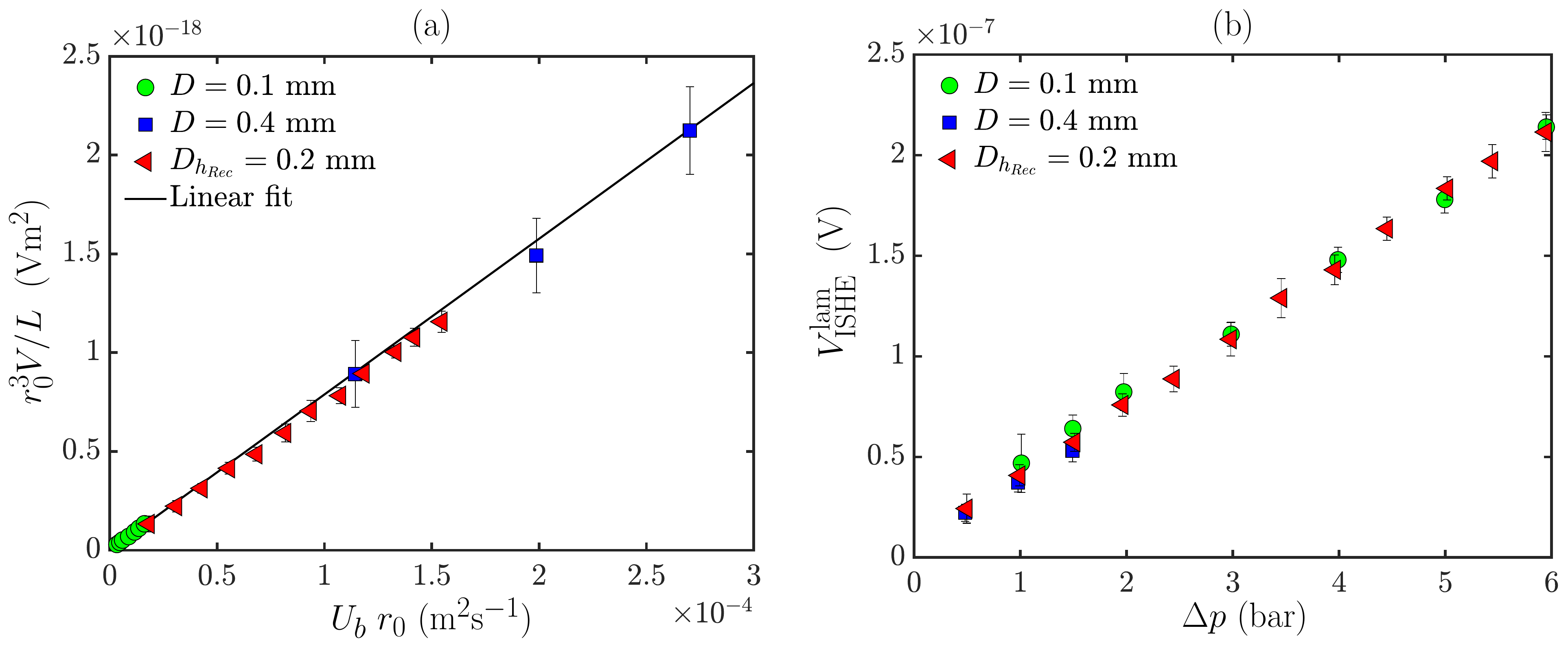}
	\caption{The voltage $V_{\mathrm{ISHE}}^{\mathrm{lam}}$ generated by laminar flows, (a) scaled based on eq.~(\ref{uni_sc_lam}) versus $(U_b~r_0)$, (b) with respect to the imposed pressure $\Delta p$ (see eq.~(\ref{pres_scale})).}
	\label{Lam_scale}
\end{figure}

Now, we investigate the validity of eq.~(\ref{uni_sc_lam}) for the generated voltage $V_{\mathrm{ISHE}}^{\mathrm{lam}}$ in laminar circular pipe flows. Figure.~\ref{Lam_scale}(a) indicates all data points scaled based on the capillary tube radius $r_0$ and length $L$ plotted versus $(U_b~r_0)$. As it can be seen, they all collapse into a single linear curve. Note that the measured voltages for the rectangular capillary tube also fall onto the same curve utilizing the hydraulic radius $r_{h0_{Rec}}=D_{h_{Rec}}/2$ as the scaling parameter. However, despite this observation, we do not argue that the hydraulic radius $r_{h0_{Rec}}$ is a universal parameter for scaling the generated voltage $V_{\mathrm{ISHE}}^{\mathrm{lam}}$ in a capillary tube with a non-circular cross section. As discussed above, to propose a universal scaling, the general solution of eq.~(\ref{main_spv}) should be provided for a capillary tube with an arbitrary cross-section. 

It is also interesting to have a look at the similarity between eq.~(\ref{uni_sc_lam}) and the well-know correlation for the pressure drop in a laminar pressure-driven flow, which reads as:
\begin{equation}
\label{pres_drop}
{\Delta p}=f_{\mathrm{lam}} \frac{L}{D} \frac{\rho U_b^{2}}{2}
\end{equation} 
Using the friction factor for a laminar flow $f_{\mathrm{lam}}={64}/{Re}$, the pressure drop correlation can be reformulated as follows:
\begin{equation}
\label{pres_scale}
\frac{r_{0}^{3}~\Delta p}{L}= 8\mu \cdot (U_b~r_0)
\end{equation} 
where $\mu$ is the dynamic viscosity of the fluid. From the similarity between eqns.(\ref{uni_sc_lam}) and (\ref{pres_scale}), it is possible to relate the generated voltage to the pressure drop of a laminar flow as follows:
\begin{equation}
\label{volt_press}
{V_{\mathrm{ISHE}}^{\mathrm{lam}}}= \frac{|e|}{\hbar} \cdot \frac{\theta_{\mathrm{SHE}} \lambda^{2} \xi_{\mathrm{lam}}}{\mu~\sigma_{0}} \cdot \Delta p
\end{equation}
\begin{figure}[h!]
	\centering
	\includegraphics[width=1\textwidth]{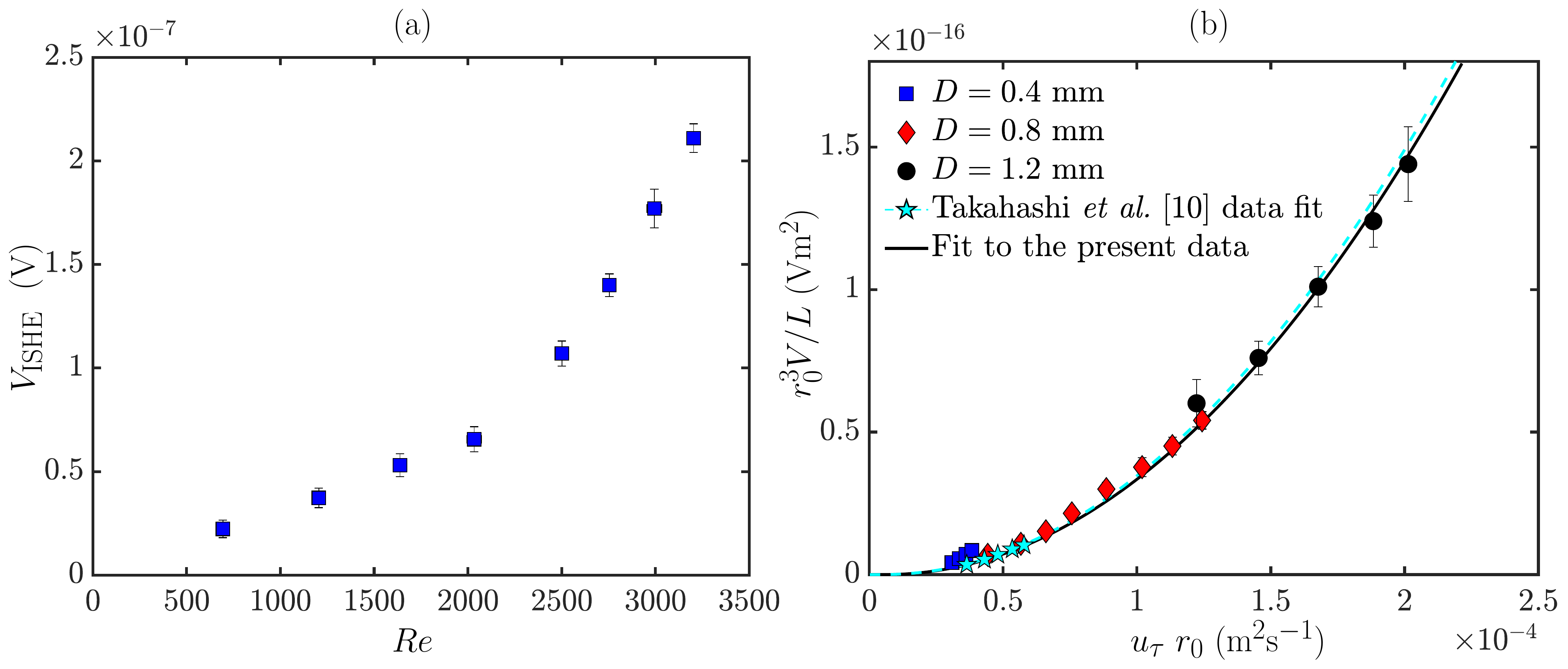}
	\caption{(a) The generated voltage $V_{\mathrm{ISHE}}$ versus Reynolds number $Re$ using a capillary tube of circular cross-section with $D=$ 0.4 mm. (b) Scaling of the generated voltage $V_{\mathrm{ISHE}}^{\mathrm{turb}}$ by turbulent flows with respect to $u_{\tau} r_0$ (see eq.~(\ref{uni_sc_turb})).}
	\label{Turb_scale}
\end{figure}
The above equation shows that using the same working liquid, the generated voltage $V_{\mathrm{ISHE}}^{\mathrm{lam}}$ should be the same at the same imposed pressure even for capillary tubes with different lengths and diameters. Figure~\ref{Lam_scale}(b) shows the generated voltages $V_{\mathrm{ISHE}}^{\mathrm{lam}}$ versus the pressure drop $\Delta p$, i.e. imposed pressure, in the circular and the rectangular capillary tubes. As it can be seen, the same voltage is obtained for all cases at the same imposed pressure. However, it should be noted that the  validity of eq.~(\ref{volt_press}) relies on the reliability of the friction factor correlation for a laminar flow $f_{\mathrm{lam}}$. While $f_{\mathrm{lam}}={64}/{Re}$ is a very well-accepted correlation for circular tubes, it has been shown that it results in a significant uncertainty (around $\pm 40\%$) in estimating the pressure drop based on the hydraulic diameter $D_h$ in non-circular tubes \cite{white2015fluid}. Nonetheless, the voltages measured here for the rectangular capillary tube show again a consistency with other data and eq.~(\ref{volt_press}). The proposed correlation in eq.~(\ref{volt_press}) may have implications for designing small-scale electric generators based on the spin hydrodynamic coupling.

\subsection{High Reynolds number turbulent flows}
Measurements were also performed for turbulent flows in circular capillary tubes with different diameters of $D= 0.4$ mm, 0.8 mm and 1.2 mm. Figure.~\ref{Turb_scale}(a) shows the measured voltage $V_{\mathrm{ISHE}}$ for a $D= 0.4$ mm capillary tube with respect to the Reynolds number $700<Re<3200$ where both laminar and turbulent regimes are covered.  Here, the transition from laminar to turbulent can be clearly identified from the measured voltages $V_{\mathrm{ISHE}}$ where it increases linearly in laminar $Re<2000$ and nonlinearly in fully turbulent $Re>2500$ regimes with the critical transition $Re$ being about 2300. As discussed before, the nonlinear behavior of the voltage with respect to the turbulent friction velocity $u_{\tau}$ which is itself a nonlinear function of the Reynolds number $Re$ was predicted and verified by \citet{takahashi2016spin} using Hg as the working liquid (see eq.~(\ref{uni_sc_turb})). These authors also reported a measurement in which they used GaInSn as the working liquid. In the present study, the data points provided in Fig. 3(b) of \citet{takahashi2016spin} are extracted and used to verify the measurements for turbulent regime. Figure.~\ref{Turb_scale} (b) shows the extracted data points (light blue pentagrams) together with our measured voltage which are scaled based on $r_{0}^3 V/L$ and plotted versus $u_{\tau} r_0$. Here, $u_{\tau}=\sqrt{\tau_{w}/\rho}$ and the wall-shear stress $\tau_{w}=C_{f} \rho U_{b}^2/2$ are estimated based on the measured bulk flow velocity $U_b$ and the well-known Blasius correlation for the friction factor $C_f=0.0791Re^{-0.25}$ for a turbulent circular pipe flow \cite{eggels1994fully}. As it can be seen from Fig.~\ref{Turb_scale}(b), the nonlinear fit to the measured data points (black solid line) based on eq.~(\ref{uni_sc_turb}) is nicely overlapped with the fit generated from the extracted data points (light blue dashed line). Indeed, the maximum Reynolds number achieved by \citet{takahashi2016spin} could be estimated based on their pipe diameter $D=0.4$ mm to be around a $Re\approx 5000$. Here, we reach a significantly higher Reynolds number of $Re=2.1\times10^4$ using a capillary tube with diameter of $D=1.2$ mm. These results show that the proposed universal scaling law by \citet{takahashi2016spin} describes very well the spin hydrodynamic phenomenon even at very high Reynolds numbers. 

We also calculated the constant coefficients of the fitted curves presented in Figs. \ref{Lam_scale}(b) and \ref{Turb_scale}(b) for the scaling laws of the spin hydrodynamic coupling for laminar and turbulent flows. Based on eqs.~(\ref{uni_sc_turb}) and ~(\ref{uni_sc_lam}), these constants can be expressed as $C_{\mathrm{lam}}=\frac{8|e|}{\hbar} \cdot \frac{(\theta_{\mathrm{SHE}} \lambda^{2} \xi_{\mathrm{lam}})}{\sigma_{0}}=7.88\times10^{-15}$ (Jsm$^{-1}$) and $C_{\mathrm{turb}}=\frac{4|e|}{\hbar} \cdot \frac{(\theta_{\mathrm{SHE}} \lambda^{2} \xi_{\mathrm{turb}})}{\sigma_{0}} \cdot \frac{1}{\kappa_+}=1.53\times10^{-14}$ (Jsm$^{-1}$) for laminar and turbulent flow, respectively. Note that except for $\xi_{\mathrm{lam}}$ and $\xi_{\mathrm{turb}}$, all the parameters in $C_{\mathrm{lam}}$ and $C_{\mathrm{turb}}$ are considered to be independent of the flow. Hence, the ratio between $\xi_{\mathrm{lam}}$ and $\xi_{\mathrm{turb}}$ can be obtained as $\xi_{\mathrm{turb}}/\xi_{\mathrm{lam}}=1.60$. 

Finally, we briefly discuss the possibility that the generated electrical voltage could be the result of a dynamo effect in the electrically conducting fluid flow. In magnetohydrodynamics, the dynamics of the magnetic field $\boldsymbol{B}$ is given by the induction equation \cite{Davidson}: 
\begin{equation}
\label{induction}
\frac{\partial \boldsymbol{B}}{\partial t}=\boldsymbol{\nabla} \times(\boldsymbol{u} \times \boldsymbol{B})+\beta \nabla^{2} \boldsymbol{B} 
\end{equation}
where $\beta$ is the magnetic diffusivity. The dimensionless magnetic Reynolds number $R_m$ relates the magnitudes of both terms on the right hand side of eq.~(\ref{induction}) to each other. In detail, $R_m=RePr_m$ where $Pr_m=\nu/\beta$ known as magnetic Prandtl number. It has been proven rigorously that for a velocity field to generate a dynamo, a certain principal rate of strain (or mean shear rate) is necessary such that the total magnetic energy $M(t)=1/\left(2 \mu_{0}\right) \int_{V} \bm{B}^{2} \mathrm{d} V$ over the whole space $V$ (with $\mu_0$ being the permeability of free space) does not decay to zero with respect to time (see \citet{Moffatt1978}). This inequality which follows from the energy balance in the magnetohydrodynamic flow can be translated into a critical magnetic Reynolds number above which a dynamo sets in, $R^\ast_m = \pi^2$ \cite{Moffatt1978}. In the present study, however, due to the low magnetic Prandtl number of GaInSn, $Pr_m=1.37\times10^{-6}$ (see Table~\ref{table:1}), the magnetic Reynolds number is limited to $R_m\lesssim 3\times 10^{-2}\ll R^\ast_m$ for the range of the flow Reynolds number $Re$ under investigation. This value is at least three orders of magnitude smaller than the well-accepted critical $R_m$ for the onset of dynamo action. The existence of a critical $R_m>10$ for triggering dynamo action in turbulent liquid metal flows is also supported by recent large scale laboratory experiments using specific flow patterns (see Refs. \cite{Gailitis2002,Gailitis2018} for reviews). Furthermore, it is believed that a non-vanishing flow helicity $H(t)=\int_V \left({\bm u}\cdot {\bm \omega}\right)\mathrm{d} V$ is required in addition for a dynamo to act \cite{Davidson}. We emphasize that a laminar flow through a straight pipe, which is shown to generate an electrical voltage for $Re \gtrsim 20$, has zero helicity. Hence, it can be concluded that the measured nano-voltage in the present study is not the result of a dynamo-generated magnetic field in combination with a classical Hall effect.

\section{Conclusions and discussion}
We have presented liquid metal pipe flow experiments in a narrow capillary that generates a measurable nano-voltage caused by the collective coupling of the spins of the freely moving electrons to the macroscopic vorticity of the charged fluid. We extended the original experiments by \citet{takahashi2016spin} in both directions with respect to the Reynolds number of the pipe flow: (1) to the laminar regime where a linear scaling of the generated electric voltage to the bulk flow velocity $U_b$ holds; (2) to higher Reynolds numbers with a fully developed turbulent circular pipe flow where the predicted scaling with respect to the turbulent friction velocity $u_{\tau}$ is found to continue to exist. We have also varied the geometry of the capillary to that of a rectangular cross-section and demonstrated the robustness of the laminar scaling. 

Our presented experiments are an independent confirmation and extension of the work of \citet{takahashi2016spin} which demonstrates the spin hydrodynamic generation without external magnetic fields, a new way of electricity generation by a coupling of spintronics with fluid dynamics. Even though the qualitative scaling behavior seems to be geometry-independent, our studies indicate the potential of an optimization of the voltage generation in several ways. Most importantly to our view, the detection of the linear dependence of the spin voltage on Reynolds number in the laminar flow regime provides an interesting starting point, as it will make the setup accessible to further microfluidic analysis at low Reynolds numbers in the future. This includes, besides analytical calculations and the variation of the cross section geometry such as different ducts and elliptical pipes, the application of nanostructured walls and/or Dean flow geometries to further maximize the vorticity gradient magnitude in the laminar flow case. The latter can result in a stronger coupling to the electron spin dynamics and thus generate a higher spin voltage without big pressure drops that are necessary for turbulent flows. These studies are currently under way and will be reported elsewhere.    

\acknowledgments
This work was financially supported by the Volkswagen Foundation. We are grateful to Christian Resagk, Till Z\"urner, Thomas Boeck, Frank Stefani and Yuri Kolesnikov for fruitful discussions, comments and remarks.
  
\bibliography{Paper}

\begin{thebibliography}{25}%
\makeatletter
\providecommand \@ifxundefined [1]{%
 \@ifx{#1\undefined}
}%
\providecommand \@ifnum [1]{%
 \ifnum #1\expandafter \@firstoftwo
 \else \expandafter \@secondoftwo
 \fi
}%
\providecommand \@ifx [1]{%
 \ifx #1\expandafter \@firstoftwo
 \else \expandafter \@secondoftwo
 \fi
}%
\providecommand \natexlab [1]{#1}%
\providecommand \enquote  [1]{``#1''}%
\providecommand \bibnamefont  [1]{#1}%
\providecommand \bibfnamefont [1]{#1}%
\providecommand \citenamefont [1]{#1}%
\providecommand \href@noop [0]{\@secondoftwo}%
\providecommand \href [0]{\begingroup \@sanitize@url \@href}%
\providecommand \@href[1]{\@@startlink{#1}\@@href}%
\providecommand \@@href[1]{\endgroup#1\@@endlink}%
\providecommand \@sanitize@url [0]{\catcode `\\12\catcode `\$12\catcode
  `\&12\catcode `\#12\catcode `\^12\catcode `\_12\catcode `\%12\relax}%
\providecommand \@@startlink[1]{}%
\providecommand \@@endlink[0]{}%
\providecommand \url  [0]{\begingroup\@sanitize@url \@url }%
\providecommand \@url [1]{\endgroup\@href {#1}{\urlprefix }}%
\providecommand \urlprefix  [0]{URL }%
\providecommand \Eprint [0]{\href }%
\providecommand \doibase [0]{https://doi.org/}%
\providecommand \selectlanguage [0]{\@gobble}%
\providecommand \bibinfo  [0]{\@secondoftwo}%
\providecommand \bibfield  [0]{\@secondoftwo}%
\providecommand \translation [1]{[#1]}%
\providecommand \BibitemOpen [0]{}%
\providecommand \bibitemStop [0]{}%
\providecommand \bibitemNoStop [0]{.\EOS\space}%
\providecommand \EOS [0]{\spacefactor3000\relax}%
\providecommand \BibitemShut  [1]{\csname bibitem#1\endcsname}%
\let\auto@bib@innerbib\@empty
\bibitem [{\citenamefont {F.~Vinen}\ and\ \citenamefont
  {J.~Donnelly}(2007)}]{Vinen2007}%
  \BibitemOpen
  \bibfield  {author} {\bibinfo {author} {\bibfnamefont {W.}~\bibnamefont
  {F.~Vinen}}\ and\ \bibinfo {author} {\bibfnamefont {R.}~\bibnamefont
  {J.~Donnelly}},\ }\bibfield  {title} {\bibinfo {title} {Quantum turbulence},\
  }\href@noop {} {\bibfield  {journal} {\bibinfo  {journal} {Phys. Today.}\
  }\textbf {\bibinfo {volume} {60}},\ \bibinfo {pages} {43} (\bibinfo {year}
  {2007})}\BibitemShut {NoStop}%
\bibitem [{\citenamefont {Fonda}\ \emph {et~al.}(2019)\citenamefont {Fonda},
  \citenamefont {Sreenivasan},\ and\ \citenamefont
  {Lathrop}}]{fonda2019reconnection}%
  \BibitemOpen
  \bibfield  {author} {\bibinfo {author} {\bibfnamefont {E.}~\bibnamefont
  {Fonda}}, \bibinfo {author} {\bibfnamefont {K.~R.}\ \bibnamefont
  {Sreenivasan}},\ and\ \bibinfo {author} {\bibfnamefont {D.~P.}\ \bibnamefont
  {Lathrop}},\ }\bibfield  {title} {\bibinfo {title} {Reconnection scaling in
  quantum fluids},\ }\href@noop {} {\bibfield  {journal} {\bibinfo  {journal}
  {Proc.Natl. Acad. Sci. USA.}\ }\textbf {\bibinfo {volume} {116}},\ \bibinfo
  {pages} {1924} (\bibinfo {year} {2019})}\BibitemShut {NoStop}%
\bibitem [{\citenamefont {Levitov}\ and\ \citenamefont
  {Falkovich}(2016)}]{levitov2016electron}%
  \BibitemOpen
  \bibfield  {author} {\bibinfo {author} {\bibfnamefont {L.}~\bibnamefont
  {Levitov}}\ and\ \bibinfo {author} {\bibfnamefont {G.}~\bibnamefont
  {Falkovich}},\ }\bibfield  {title} {\bibinfo {title} {Electron viscosity,
  current vortices and negative nonlocal resistance in graphene},\ }\href@noop
  {} {\bibfield  {journal} {\bibinfo  {journal} {Nat. Phys.}\ }\textbf
  {\bibinfo {volume} {12}},\ \bibinfo {pages} {672} (\bibinfo {year}
  {2016})}\BibitemShut {NoStop}%
\bibitem [{\citenamefont {Wolf}\ \emph {et~al.}(2001)\citenamefont {Wolf},
  \citenamefont {Awschalom}, \citenamefont {Buhrman}, \citenamefont {Daughton},
  \citenamefont {Von~Molnar}, \citenamefont {Roukes}, \citenamefont
  {Chtchelkanova},\ and\ \citenamefont {Treger}}]{wolf2001spintronics}%
  \BibitemOpen
  \bibfield  {author} {\bibinfo {author} {\bibfnamefont {S.}~\bibnamefont
  {Wolf}}, \bibinfo {author} {\bibfnamefont {D.}~\bibnamefont {Awschalom}},
  \bibinfo {author} {\bibfnamefont {R.}~\bibnamefont {Buhrman}}, \bibinfo
  {author} {\bibfnamefont {J.}~\bibnamefont {Daughton}}, \bibinfo {author}
  {\bibfnamefont {S.}~\bibnamefont {Von~Molnar}}, \bibinfo {author}
  {\bibfnamefont {M.}~\bibnamefont {Roukes}}, \bibinfo {author} {\bibfnamefont
  {A.~Y.}\ \bibnamefont {Chtchelkanova}},\ and\ \bibinfo {author}
  {\bibfnamefont {D.}~\bibnamefont {Treger}},\ }\bibfield  {title} {\bibinfo
  {title} {Spintronics: a spin-based electronics vision for the future},\
  }\href@noop {} {\bibfield  {journal} {\bibinfo  {journal} {Science}\ }\textbf
  {\bibinfo {volume} {294}},\ \bibinfo {pages} {1488} (\bibinfo {year}
  {2001})}\BibitemShut {NoStop}%
\bibitem [{\citenamefont {{\v{Z}}uti{\'c}}\ \emph {et~al.}(2004)\citenamefont
  {{\v{Z}}uti{\'c}}, \citenamefont {Fabian},\ and\ \citenamefont
  {Sarma}}]{vzutic2004spintronics}%
  \BibitemOpen
  \bibfield  {author} {\bibinfo {author} {\bibfnamefont {I.}~\bibnamefont
  {{\v{Z}}uti{\'c}}}, \bibinfo {author} {\bibfnamefont {J.}~\bibnamefont
  {Fabian}},\ and\ \bibinfo {author} {\bibfnamefont {S.~D.}\ \bibnamefont
  {Sarma}},\ }\bibfield  {title} {\bibinfo {title} {Spintronics: Fundamentals
  and applications},\ }\href@noop {} {\bibfield  {journal} {\bibinfo  {journal}
  {Rev. Mod. Phys.}\ }\textbf {\bibinfo {volume} {76}},\ \bibinfo {pages} {323}
  (\bibinfo {year} {2004})}\BibitemShut {NoStop}%
\bibitem [{\citenamefont {Hirohata}\ and\ \citenamefont
  {Takanashi}(2014)}]{hirohata2014future}%
  \BibitemOpen
  \bibfield  {author} {\bibinfo {author} {\bibfnamefont {A.}~\bibnamefont
  {Hirohata}}\ and\ \bibinfo {author} {\bibfnamefont {K.}~\bibnamefont
  {Takanashi}},\ }\bibfield  {title} {\bibinfo {title} {Future perspectives for
  spintronic devices},\ }\href@noop {} {\bibfield  {journal} {\bibinfo
  {journal} {J. Phys. D. Appl. Phys.}\ }\textbf {\bibinfo {volume} {47}},\
  \bibinfo {pages} {193001} (\bibinfo {year} {2014})}\BibitemShut {NoStop}%
\bibitem [{\citenamefont {Ando}\ \emph {et~al.}(2010)\citenamefont {Ando},
  \citenamefont {Morikawa}, \citenamefont {Trypiniotis}, \citenamefont
  {Fujikawa}, \citenamefont {Barnes},\ and\ \citenamefont
  {Saitoh}}]{ando2010direct}%
  \BibitemOpen
  \bibfield  {author} {\bibinfo {author} {\bibfnamefont {K.}~\bibnamefont
  {Ando}}, \bibinfo {author} {\bibfnamefont {M.}~\bibnamefont {Morikawa}},
  \bibinfo {author} {\bibfnamefont {T.}~\bibnamefont {Trypiniotis}}, \bibinfo
  {author} {\bibfnamefont {Y.}~\bibnamefont {Fujikawa}}, \bibinfo {author}
  {\bibfnamefont {C.}~\bibnamefont {Barnes}},\ and\ \bibinfo {author}
  {\bibfnamefont {E.}~\bibnamefont {Saitoh}},\ }\bibfield  {title} {\bibinfo
  {title} {Direct conversion of light-polarization information into electric
  voltage using photoinduced inverse spin-hall effect in pt/gaas hybrid
  structure: Spin photodetector},\ }\href@noop {} {\bibfield  {journal}
  {\bibinfo  {journal} {J. Appl. Phys.}\ }\textbf {\bibinfo {volume} {107}},\
  \bibinfo {pages} {113902} (\bibinfo {year} {2010})}\BibitemShut {NoStop}%
\bibitem [{\citenamefont {Sandweg}\ \emph {et~al.}(2011)\citenamefont
  {Sandweg}, \citenamefont {Kajiwara}, \citenamefont {Chumak}, \citenamefont
  {Serga}, \citenamefont {Vasyuchka}, \citenamefont {Jungfleisch},
  \citenamefont {Saitoh},\ and\ \citenamefont {Hillebrands}}]{sandweg2011spin}%
  \BibitemOpen
  \bibfield  {author} {\bibinfo {author} {\bibfnamefont {C.~W.}\ \bibnamefont
  {Sandweg}}, \bibinfo {author} {\bibfnamefont {Y.}~\bibnamefont {Kajiwara}},
  \bibinfo {author} {\bibfnamefont {A.~V.}\ \bibnamefont {Chumak}}, \bibinfo
  {author} {\bibfnamefont {A.~A.}\ \bibnamefont {Serga}}, \bibinfo {author}
  {\bibfnamefont {V.~I.}\ \bibnamefont {Vasyuchka}}, \bibinfo {author}
  {\bibfnamefont {M.~B.}\ \bibnamefont {Jungfleisch}}, \bibinfo {author}
  {\bibfnamefont {E.}~\bibnamefont {Saitoh}},\ and\ \bibinfo {author}
  {\bibfnamefont {B.}~\bibnamefont {Hillebrands}},\ }\bibfield  {title}
  {\bibinfo {title} {Spin pumping by parametrically excited exchange magnons},\
  }\href@noop {} {\bibfield  {journal} {\bibinfo  {journal} {Phys. Rev. Lett.}\
  }\textbf {\bibinfo {volume} {106}},\ \bibinfo {pages} {216601} (\bibinfo
  {year} {2011})}\BibitemShut {NoStop}%
\bibitem [{\citenamefont {Uchida}\ \emph {et~al.}(2008)\citenamefont {Uchida},
  \citenamefont {Takahashi}, \citenamefont {Harii}, \citenamefont {Ieda},
  \citenamefont {Koshibae}, \citenamefont {Ando}, \citenamefont {Maekawa},\
  and\ \citenamefont {Saitoh}}]{uchida2008observation}%
  \BibitemOpen
  \bibfield  {author} {\bibinfo {author} {\bibfnamefont {K.}~\bibnamefont
  {Uchida}}, \bibinfo {author} {\bibfnamefont {S.}~\bibnamefont {Takahashi}},
  \bibinfo {author} {\bibfnamefont {K.}~\bibnamefont {Harii}}, \bibinfo
  {author} {\bibfnamefont {J.}~\bibnamefont {Ieda}}, \bibinfo {author}
  {\bibfnamefont {W.}~\bibnamefont {Koshibae}}, \bibinfo {author}
  {\bibfnamefont {K.}~\bibnamefont {Ando}}, \bibinfo {author} {\bibfnamefont
  {S.}~\bibnamefont {Maekawa}},\ and\ \bibinfo {author} {\bibfnamefont
  {E.}~\bibnamefont {Saitoh}},\ }\bibfield  {title} {\bibinfo {title}
  {Observation of the spin seebeck effect},\ }\href@noop {} {\bibfield
  {journal} {\bibinfo  {journal} {Nature}\ }\textbf {\bibinfo {volume} {455}},\
  \bibinfo {pages} {778} (\bibinfo {year} {2008})}\BibitemShut {NoStop}%
\bibitem [{\citenamefont {Takahashi}\ \emph {et~al.}(2016)\citenamefont
  {Takahashi}, \citenamefont {Matsuo}, \citenamefont {Ono}, \citenamefont
  {Harii}, \citenamefont {Chudo}, \citenamefont {Okayasu}, \citenamefont
  {Ieda}, \citenamefont {Takahashi}, \citenamefont {Maekawa},\ and\
  \citenamefont {Saitoh}}]{takahashi2016spin}%
  \BibitemOpen
  \bibfield  {author} {\bibinfo {author} {\bibfnamefont {R.}~\bibnamefont
  {Takahashi}}, \bibinfo {author} {\bibfnamefont {M.}~\bibnamefont {Matsuo}},
  \bibinfo {author} {\bibfnamefont {M.}~\bibnamefont {Ono}}, \bibinfo {author}
  {\bibfnamefont {K.}~\bibnamefont {Harii}}, \bibinfo {author} {\bibfnamefont
  {H.}~\bibnamefont {Chudo}}, \bibinfo {author} {\bibfnamefont
  {S.}~\bibnamefont {Okayasu}}, \bibinfo {author} {\bibfnamefont
  {J.}~\bibnamefont {Ieda}}, \bibinfo {author} {\bibfnamefont {S.}~\bibnamefont
  {Takahashi}}, \bibinfo {author} {\bibfnamefont {S.}~\bibnamefont {Maekawa}},\
  and\ \bibinfo {author} {\bibfnamefont {E.}~\bibnamefont {Saitoh}},\
  }\bibfield  {title} {\bibinfo {title} {Spin hydrodynamic generation},\
  }\href@noop {} {\bibfield  {journal} {\bibinfo  {journal} {Nat. Phys.}\
  }\textbf {\bibinfo {volume} {12}},\ \bibinfo {pages} {52} (\bibinfo {year}
  {2016})}\BibitemShut {NoStop}%
\bibitem [{\citenamefont {Matsuo}\ \emph {et~al.}(2013)\citenamefont {Matsuo},
  \citenamefont {Ieda}, \citenamefont {Harii}, \citenamefont {Saitoh},\ and\
  \citenamefont {Maekawa}}]{matsuo2013mechanical}%
  \BibitemOpen
  \bibfield  {author} {\bibinfo {author} {\bibfnamefont {M.}~\bibnamefont
  {Matsuo}}, \bibinfo {author} {\bibfnamefont {J.}~\bibnamefont {Ieda}},
  \bibinfo {author} {\bibfnamefont {K.}~\bibnamefont {Harii}}, \bibinfo
  {author} {\bibfnamefont {E.}~\bibnamefont {Saitoh}},\ and\ \bibinfo {author}
  {\bibfnamefont {S.}~\bibnamefont {Maekawa}},\ }\bibfield  {title} {\bibinfo
  {title} {Mechanical generation of spin current by spin-rotation coupling},\
  }\href@noop {} {\bibfield  {journal} {\bibinfo  {journal} {Phys. Rev. B.}\
  }\textbf {\bibinfo {volume} {87}},\ \bibinfo {pages} {180402} (\bibinfo
  {year} {2013})}\BibitemShut {NoStop}%
\bibitem [{\citenamefont {Sinova}\ \emph {et~al.}(2015)\citenamefont {Sinova},
  \citenamefont {Valenzuela}, \citenamefont {Wunderlich}, \citenamefont
  {Back},\ and\ \citenamefont {Jungwirth}}]{sinova2015spin}%
  \BibitemOpen
  \bibfield  {author} {\bibinfo {author} {\bibfnamefont {J.}~\bibnamefont
  {Sinova}}, \bibinfo {author} {\bibfnamefont {S.~O.}\ \bibnamefont
  {Valenzuela}}, \bibinfo {author} {\bibfnamefont {J.}~\bibnamefont
  {Wunderlich}}, \bibinfo {author} {\bibfnamefont {C.}~\bibnamefont {Back}},\
  and\ \bibinfo {author} {\bibfnamefont {T.}~\bibnamefont {Jungwirth}},\
  }\bibfield  {title} {\bibinfo {title} {Spin hall effects},\ }\href@noop {}
  {\bibfield  {journal} {\bibinfo  {journal} {Rev. Mod. Phys.}\ }\textbf
  {\bibinfo {volume} {87}},\ \bibinfo {pages} {1213} (\bibinfo {year}
  {2015})}\BibitemShut {NoStop}%
\bibitem [{\citenamefont {Saitoh}\ \emph {et~al.}(2006)\citenamefont {Saitoh},
  \citenamefont {Ueda}, \citenamefont {Miyajima},\ and\ \citenamefont
  {Tatara}}]{saitoh2006conversion}%
  \BibitemOpen
  \bibfield  {author} {\bibinfo {author} {\bibfnamefont {E.}~\bibnamefont
  {Saitoh}}, \bibinfo {author} {\bibfnamefont {M.}~\bibnamefont {Ueda}},
  \bibinfo {author} {\bibfnamefont {H.}~\bibnamefont {Miyajima}},\ and\
  \bibinfo {author} {\bibfnamefont {G.}~\bibnamefont {Tatara}},\ }\bibfield
  {title} {\bibinfo {title} {Conversion of spin current into charge current at
  room temperature: Inverse spin-hall effect},\ }\href@noop {} {\bibfield
  {journal} {\bibinfo  {journal} {App. Phys. lett.}\ }\textbf {\bibinfo
  {volume} {88}},\ \bibinfo {pages} {182509} (\bibinfo {year}
  {2006})}\BibitemShut {NoStop}%
\bibitem [{\citenamefont {White}()}]{white2015fluid}%
  \BibitemOpen
  \bibfield  {author} {\bibinfo {author} {\bibfnamefont {F.~M.}\ \bibnamefont
  {White}},\ }\bibfield  {title} {\bibinfo {title} {\textit{Fluid mechanics}},\
  }\href@noop {} {\bibinfo  {journal} {(McGraw-Hill, New York, 2015)}\
  }\BibitemShut {NoStop}%
\bibitem [{\citenamefont {Matsuo}\ \emph {et~al.}(2017)\citenamefont {Matsuo},
  \citenamefont {Ohnuma},\ and\ \citenamefont {Maekawa}}]{matsuo2017theory}%
  \BibitemOpen
\bibfield  {journal} {  }\bibfield  {author} {\bibinfo {author} {\bibfnamefont
  {M.}~\bibnamefont {Matsuo}}, \bibinfo {author} {\bibfnamefont
  {Y.}~\bibnamefont {Ohnuma}},\ and\ \bibinfo {author} {\bibfnamefont
  {S.}~\bibnamefont {Maekawa}},\ }\bibfield  {title} {\bibinfo {title} {Theory
  of spin hydrodynamic generation},\ }\href@noop {} {\bibfield  {journal}
  {\bibinfo  {journal} {Phys. Rev. B.}\ }\textbf {\bibinfo {volume} {96}},\
  \bibinfo {pages} {020401} (\bibinfo {year} {2017})}\BibitemShut {NoStop}%
\bibitem [{\citenamefont {Plevachuk}\ \emph {et~al.}(2014)\citenamefont
  {Plevachuk}, \citenamefont {Sklyarchuk}, \citenamefont {Eckert},
  \citenamefont {Gerbeth},\ and\ \citenamefont {Novakovic}}]{plevachuk2014}%
  \BibitemOpen
  \bibfield  {author} {\bibinfo {author} {\bibfnamefont {Y.}~\bibnamefont
  {Plevachuk}}, \bibinfo {author} {\bibfnamefont {V.}~\bibnamefont
  {Sklyarchuk}}, \bibinfo {author} {\bibfnamefont {S.}~\bibnamefont {Eckert}},
  \bibinfo {author} {\bibfnamefont {G.}~\bibnamefont {Gerbeth}},\ and\ \bibinfo
  {author} {\bibfnamefont {R.}~\bibnamefont {Novakovic}},\ }\bibfield  {title}
  {\bibinfo {title} {Thermophysical properties of the liquid ga-in-sn eutectic
  alloy},\ }\href@noop {} {\bibfield  {journal} {\bibinfo  {journal} {J. Chem.
  Eng. Data}\ }\textbf {\bibinfo {volume} {59}},\ \bibinfo {pages} {757}
  (\bibinfo {year} {2014})}\BibitemShut {NoStop}%
\bibitem [{\citenamefont {Eckert}(2020)}]{Sven}%
  \BibitemOpen
  \bibfield  {author} {\bibinfo {author} {\bibfnamefont {S.}~\bibnamefont
  {Eckert}},\ }\bibfield  {title} {\bibinfo {title} {private communication},\
  }\href@noop {} {\  (\bibinfo {year} {2020})}\BibitemShut {NoStop}%
\bibitem [{\citenamefont {Valet}\ and\ \citenamefont
  {Fert}(1993)}]{valet1993theory}%
  \BibitemOpen
  \bibfield  {author} {\bibinfo {author} {\bibfnamefont {T.}~\bibnamefont
  {Valet}}\ and\ \bibinfo {author} {\bibfnamefont {A.}~\bibnamefont {Fert}},\
  }\bibfield  {title} {\bibinfo {title} {Theory of the perpendicular
  magnetoresistance in magnetic multilayers},\ }\href@noop {} {\bibfield
  {journal} {\bibinfo  {journal} {Phys. Rev. B.}\ }\textbf {\bibinfo {volume}
  {48}},\ \bibinfo {pages} {7099} (\bibinfo {year} {1993})}\BibitemShut
  {NoStop}%
\bibitem [{\citenamefont {Pope}()}]{pope2001turbulent}%
  \BibitemOpen
  \bibfield  {author} {\bibinfo {author} {\bibfnamefont {S.~B.}\ \bibnamefont
  {Pope}},\ }\bibfield  {title} {\bibinfo {title} {\textit{Turbulent flows}},\
  }\href@noop {} {\bibinfo  {journal} {(Cambridge University Press, Cambirdge,
  United Kingdom, 2001)}\ }\BibitemShut {NoStop}%
\bibitem [{\citenamefont {Takahashi}\ and\ \citenamefont
  {Maekawa}(2008)}]{takahashi2008spin}%
  \BibitemOpen
\bibfield  {journal} {  }\bibfield  {author} {\bibinfo {author} {\bibfnamefont
  {S.}~\bibnamefont {Takahashi}}\ and\ \bibinfo {author} {\bibfnamefont
  {S.}~\bibnamefont {Maekawa}},\ }\bibfield  {title} {\bibinfo {title} {Spin
  current, spin accumulation and spin hall effect},\ }\href@noop {} {\bibfield
  {journal} {\bibinfo  {journal} {Sci. Technol. Adv. Mat.}\ }\textbf {\bibinfo
  {volume} {9}},\ \bibinfo {pages} {014105} (\bibinfo {year}
  {2008})}\BibitemShut {NoStop}%
\bibitem [{\citenamefont {Eggels}\ \emph {et~al.}(1994)\citenamefont {Eggels},
  \citenamefont {Unger}, \citenamefont {Weiss}, \citenamefont {Westerweel},
  \citenamefont {Adrian}, \citenamefont {Friedrich},\ and\ \citenamefont
  {Nieuwstadt}}]{eggels1994fully}%
  \BibitemOpen
  \bibfield  {author} {\bibinfo {author} {\bibfnamefont {J.}~\bibnamefont
  {Eggels}}, \bibinfo {author} {\bibfnamefont {F.}~\bibnamefont {Unger}},
  \bibinfo {author} {\bibfnamefont {M.}~\bibnamefont {Weiss}}, \bibinfo
  {author} {\bibfnamefont {J.}~\bibnamefont {Westerweel}}, \bibinfo {author}
  {\bibfnamefont {R.}~\bibnamefont {Adrian}}, \bibinfo {author} {\bibfnamefont
  {R.}~\bibnamefont {Friedrich}},\ and\ \bibinfo {author} {\bibfnamefont
  {F.}~\bibnamefont {Nieuwstadt}},\ }\bibfield  {title} {\bibinfo {title}
  {Fully developed turbulent pipe flow: a comparison between direct numerical
  simulation and experiment},\ }\href@noop {} {\bibfield  {journal} {\bibinfo
  {journal} {J. Fluid. Mech.}\ }\textbf {\bibinfo {volume} {268}},\ \bibinfo
  {pages} {175} (\bibinfo {year} {1994})}\BibitemShut {NoStop}%
\bibitem [{\citenamefont {Davidson}()}]{Davidson}%
  \BibitemOpen
  \bibfield  {author} {\bibinfo {author} {\bibfnamefont {P.~A.}\ \bibnamefont
  {Davidson}},\ }\bibfield  {title} {\bibinfo {title} {\textit{An Introduction
  to Magnetohydrodynamics}},\ }\href@noop {} {\bibinfo  {journal} {(Cambridge
  University Press, Cambridge, United Kingdom, 2001)}\ }\BibitemShut {NoStop}%
\bibitem [{\citenamefont {Moffatt}()}]{Moffatt1978}%
  \BibitemOpen
\bibfield  {journal} {  }\bibfield  {author} {\bibinfo {author} {\bibfnamefont
  {H.~K.}\ \bibnamefont {Moffatt}},\ }\bibfield  {title} {\bibinfo {title}
  {\textit{Magnetic field generation in electrically conducting fluids}},\
  }\href@noop {} {\bibinfo  {journal} {(Cambridge University Press, Cambridge,
  United Kingdom, 1978)}\ }\BibitemShut {NoStop}%
\bibitem [{\citenamefont {Gailitis}\ \emph {et~al.}(2002)\citenamefont
  {Gailitis}, \citenamefont {Lielausis}, \citenamefont {Platacis},
  \citenamefont {Gerbeth},\ and\ \citenamefont {Stefani}}]{Gailitis2002}%
  \BibitemOpen
\bibfield  {journal} {  }\bibfield  {author} {\bibinfo {author} {\bibfnamefont
  {A.}~\bibnamefont {Gailitis}}, \bibinfo {author} {\bibfnamefont
  {O.}~\bibnamefont {Lielausis}}, \bibinfo {author} {\bibfnamefont
  {E.}~\bibnamefont {Platacis}}, \bibinfo {author} {\bibfnamefont
  {G.}~\bibnamefont {Gerbeth}},\ and\ \bibinfo {author} {\bibfnamefont
  {F.}~\bibnamefont {Stefani}},\ }\bibfield  {title} {\bibinfo {title}
  {Colloquium: Laboratory experiments in hydromagnetic dynamos},\ }\href@noop
  {} {\bibfield  {journal} {\bibinfo  {journal} {Rev. Mod. Phys.}\ }\textbf
  {\bibinfo {volume} {74}},\ \bibinfo {pages} {973} (\bibinfo {year}
  {2002})}\BibitemShut {NoStop}%
\bibitem [{\citenamefont {Gailitis}\ \emph {et~al.}(2018)\citenamefont
  {Gailitis}, \citenamefont {Gerbeth}, \citenamefont {Gundrum}, \citenamefont
  {Lielausis}, \citenamefont {Lipsbergs}, \citenamefont {Platacis},\ and\
  \citenamefont {Stefani}}]{Gailitis2018}%
  \BibitemOpen
  \bibfield  {author} {\bibinfo {author} {\bibfnamefont {A.}~\bibnamefont
  {Gailitis}}, \bibinfo {author} {\bibfnamefont {G.}~\bibnamefont {Gerbeth}},
  \bibinfo {author} {\bibfnamefont {T.}~\bibnamefont {Gundrum}}, \bibinfo
  {author} {\bibfnamefont {O.}~\bibnamefont {Lielausis}}, \bibinfo {author}
  {\bibfnamefont {G.}~\bibnamefont {Lipsbergs}}, \bibinfo {author}
  {\bibfnamefont {E.}~\bibnamefont {Platacis}},\ and\ \bibinfo {author}
  {\bibfnamefont {F.}~\bibnamefont {Stefani}},\ }\bibfield  {title} {\bibinfo
  {title} {Self-excitation in a helical liquid metal flow: the riga dynamo
  experiments},\ }\href@noop {} {\bibfield  {journal} {\bibinfo  {journal} {J.
  Plasma Phys.}\ }\textbf {\bibinfo {volume} {84}},\ \bibinfo {pages}
  {735840301} (\bibinfo {year} {2018})}\BibitemShut {NoStop}%
\end{thebibliography}%

\end{document}